\documentclass{optica-article}

\journal{opticajournal} 

\articletype{Research Article}

\usepackage{lineno, xspace, float}

\newcommand{\Poincare}{Poincar\'e\xspace}

\begin{document}

\title{A wavefront rotator with near-zero mean polarization change}

\author{Suman Karan,\authormark{1,*} Nilakshi Senapati,\authormark{1} and Anand K. Jha\authormark{1,*}}

\address{\authormark{1}Department of Physics, Indian Institute of Technology Kanpur, Kanpur, Uttar Pradesh 208016, India}

\email{\authormark{*}karans@iitk.ac.in\\
\authormark{*}akjha@iitk.ac.in} 


\begin{abstract*} 
A K-mirror is a device that rotates the wavefront of an incident optical field. It has recently gained prominence over Dove prism, another commonly used wavefront rotator, due to the fact that while a K-mirror has several controls for adjusting the internal reflections, a Dove prism is made of a single glass element with no additional control. Thus, one can obtain much lower angular deviations of transmitting wavefronts using a K-mirror than with a Dove prism. However, the accompanying polarization changes in the transmitted field due to rotation persist even in the commercially available K-mirrors. A recent theoretical work [Applied Optics, 61, 8302 (2022)] shows that it is possible to optimize the base angle of a K-mirror for a given refractive index  such that the accompanying polarization changes are minimum. In contrast, we show in this article that by optimizing the refractive index it is possible to design a K-mirror at any given base angle and with any given value for the mean polarization change, including near-zero values. Furthermore, we experimentally demonstrate a K-mirror with an order-of-magnitude lower mean polarization change than that of the commercially available K-mirrors. This can have important practical implications for OAM-based  applications that require precise wavefront rotation control. 

\end{abstract*}


\section{Introduction}

Wavefront rotation plays a crucial role in measuring the orbital angular momentum (OAM) of optical fields \cite{wang2017oe, kulkarni2020prapplied,  pires2010ol} as well of single-photon fields \cite{leach2002prl,leach2004prl,pires2010prl}. Wavefront rotators are also indispensable for many other applications, including interferometry \cite{moreno2003ao,chu2008oe,mohanty1983ocomm}, beam steering \cite{pantsar2009ical}, microscopy \cite{zhi2015ol}, optical astronomy \cite{wildey1967asp}, pattern recognition \cite{fujii1981ocomm}, and holography \cite{lopez2001ao}. Although there exists a variety of wavefront rotators \cite{padgett1999jomo, moreno200ao, leach2004prl, liu2009josaa}, the Dove prism \cite{padgett1999jomo} and the K-mirror \cite{leach2004prl} are the most commonly used. 

A Dove prism is a very easy-to-use commercially-available wavefront rotator and is made of a single glass element without many controls. The absence of controls results in the angular deviations of fields through the commercially available Dove prisms being in tens of milli-radians, which makes them unsuitable for precision experiments \cite{pires2010ol, pires2010prl}. On the other hand, a K-mirror consists of three mirrors with independent controls, which can be used for better alignment capabilities and thus for minimizing the angular deviations to less than $200$ $\mu$-${\rm radian}$ \cite{peeters2007pra}. Consequently, K-mirrors have found use in vibrations measurement of rotation objects \cite{altmann2016VibrationMO, rohloff2014oengg}, telescopic tracking  \cite{buckle2009mnrs, guo2014symposium}, spatial coherence measurement \cite{pires2010josaa}, and OAM spectrum measurement   \cite{pires2010prl, pires2010ol}. Nonetheless, both K-mirrors and Dove prisms introduce polarization changes in the transmitting field as a function of rotation, which limits their range of applicability. A recent theoretical work has shown that it is possible to optimize the base angle of a K-mirror for a given refractive index  such that the accompanying polarization changes are minimum \cite{karan2022ao}. However, no experimental demonstration of such a K-mirror has been reported so far. 

In this article, we generalize the result reported in Ref.~\cite{karan2022ao}. Using an optimization technique that is independent of the input polarization, we show that it is possible to design a K-mirror at any given base angle and with any given value for the mean polarization change. Furthermore, we experimentally demonstrate a K-mirror with an order-of-magnitude lower mean polarization changes than that of the commercially available K-mirrors.



%
%
\begin{figure*}[t!]
\centering
\includegraphics[scale=0.86]{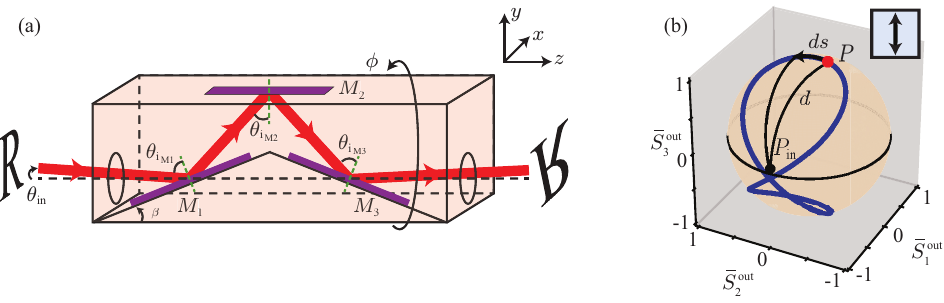}
\caption{(a) Schematic of the wavefront rotator, a K-mirror with base angle $\beta$. (b) \Poincare sphere representation of the transmitted field as a function of the rotation angle $\phi$. The black dot $P_{\rm in}$ represents the input state of polarization, also shown in the inset. The red dot $P$ denotes the transmitted state of polarization at rotation angle $\phi$, which ranges from $0^{\circ}$ to $180^{\circ}$. The distance $d$ is the geodesic distance between points $P$ and $P_{\rm in}$ on the surface of the \Poincare sphere, and $ds$ is the infinitesimal small arc length. }
\label{fig:schematic_of_km_and_poincare}
\end{figure*}

\section{Theory}
This section is mostly based on the theoretical treatment given in Ref.~\cite{karan2022ao}. When an incident electric field ${\bf E}^{\rm in} $ goes through a K-mirror with base angle $ \beta$, it undergoes three mirror reflections as depicted in Fig.~\ref{fig:schematic_of_km_and_poincare}(a). Now, for ${\bf E}^{\rm in}= E_{x}^{\rm in} \hat{\bf x}+ E_{y}^{\rm in} \hat{\bf y}$, where $ E_{x}^{\rm in}=  \cos \psi_{\rm in}$ and $ E_{y}^{\rm in}=\sin \psi_{\rm in} e^{i \delta_{\rm in}}$ are the incident electric field component along $x-$ and $y-$ polarized directions, the incident state of polarization can be expressed in terms of four Stokes parameters: $S^{\rm in}_0 = | E_{x}^{\rm in}|^2 + | E_{y}^{\rm in}|^2 $, $S^{\rm in}_1 =  | E_{x}^{\rm in}|^2 - | E_{y}^{\rm in}|^2 $, $S^{\rm in}_2 = 2~ {\rm Re} \left[ E_{x}^{{\rm in}*} E_{y}^{\rm in} \right] $, and $S^{\rm in}_3 = 2~ {\rm Im} \left[ E_{x}^{{\rm in}*} E_{y}^{\rm in} \right]$ \cite{goldstein2017crc}. We can normalize the Stokes parameters and represent them as $\bar{S}^{\rm in}_0 = S^{\rm in}_0 / S^{\rm in}_0$, $\bar{S}^{\rm in}_1 = S^{\rm in}_1 / S^{\rm in}_0$, $\bar{S}^{\rm in}_2 = S^{\rm in}_2 / S^{\rm in}_0$, and $\bar{S}^{\rm in}_3 = S^{\rm in}_3 / S^{\rm in}_0$. Therefore, the normalized Stokes parameters can be graphically represented as a point on the surface of the \Poincare sphere. Figure~\ref{fig:schematic_of_km_and_poincare}(b) depicts schematic of the \Poincare sphere, where the black dot $P_{\rm in}$ represents the vertically polarized incident state. The transmitted field ${\bf E}^{\rm out} =  E_{x}^{\rm out} \hat{\bf x}+ E_{y}^{\rm out} \hat{\bf y}$ at any arbitrary rotation angle of the K-mirror $\phi$ can be written as ${\bf E}^{\rm out} = T_{\rm KM}\left(\phi\right) {\bf E}^{\rm in}$, where $T_{\rm KM}\left(\phi\right)$ is the transfer matrix for K-mirror at any arbitrary rotation angle $\phi$ and it can be expressed as \cite{karan2022ao}
\begin{align}
\label{eqn: transfer_mtrix}
 T_{\rm KM}\left(\phi\right) 
   =
   \begin{bmatrix}
    \cos \phi & - \sin \phi \\
    \sin \phi & \cos \phi
    \end{bmatrix}
   \begin{bmatrix}
        T^{s}_{\rm KM} & 0\\
        0 &  T^{p}_{\rm KM}
       \end{bmatrix}
        \begin{bmatrix}
    \cos \phi & - \sin \phi \\
    \sin \phi & \cos \phi
    \end{bmatrix}^{T},     
\end{align}
where 
\begin{align}
 T^{\rm s}_{\rm KM}=  - \dfrac{\sin(\theta_{\rm i_{M1}} -~ \theta_{\rm t_{M1}} ) }{\sin(\theta_{\rm i_{M1}} +~ \theta_{\rm t_{M1}} )}\times
 \dfrac{\sin(\theta_{\rm i_{M2}} -~ \theta_{\rm t_{M2}} ) }{\sin(\theta_{\rm i_{M2}} +~ \theta_{\rm t_{M2}} )} \times \dfrac{\sin(\theta_{\rm i_{M3}} -~ \theta_{\rm t_{M3}} ) }{\sin(\theta_{\rm i_{M3}} +~ \theta_{\rm t_{M3}} )},
\end{align}
and
\begin{align}
 T^{\rm p}_{\rm KM}= \dfrac{\tan(\theta_{\rm i_{M1}} -~ \theta_{\rm t_{M1}} ) }{\tan(\theta_{\rm i_{M1}} +~ \theta_{\rm t_{M1}} )}\times \dfrac{\tan(\theta_{\rm i_{M2}} -~ \theta_{\rm t_{M2}} ) }{\tan(\theta_{\rm i_{M2}} +~ \theta_{\rm t_{M2}} )}\times \dfrac{\tan(\theta_{\rm i_{M3}} -~ \theta_{\rm t_{M3}} ) }{\tan(\theta_{\rm i_{M3}} +~ \theta_{\rm t_{M3}} )}.
\end{align}
Here $\theta_{\rm i_{M1}} = \pi/2- (\beta +\theta_{\rm in})$,  $\theta_{\rm i_{M2}} = \pi/2- (2\beta +\theta_{\rm in})$,  and $\theta_{\rm i_{M3}} = \pi/2- (\beta +\theta_{\rm in})$ are the angles of incidence on mirrors $M_1$, $M_2$, and $M_3$, respectively, as shown in Fig.~\ref{fig:schematic_of_km_and_poincare}(a). The corresponding angles of transmission are given by $\theta_{\rm t_{M1}}$, $\theta_{\rm t_{M2}}$, and $\theta_{\rm t_{M3}}$, and these can be written as $n_M \sin \theta_{\rm t_{M1}} = \sin \theta_{\rm i_{M1}}$, 
$n_M\sin \theta_{\rm t_{M2}} = \sin \theta_{\rm i_{M2}}$, and  $n_M \sin \theta_{\rm t_{M3}} =  \sin \theta_{\rm i_{M3}}$. Here $n_M$ is the refractive index of the reflective coating of the mirror. $\beta$ is the base angle and $\theta_{\rm in}$ is the angle between the propagation direction and the rotation axis. We note that the details of how the Fresnel coefficients for each surface is calculated has been worked out in Ref.~\cite{karan2022ao}. In this context, we further note that since we are considering a general reflective coating including metallic coating, we take $n_M$ to be a complex quantity. As a consequence, the Fresnel coefficients as well as the transmission angles, which are needed for calculating the Fresnel coefficients, are in general complex quantities.

The Stokes parameters of the transmitted field ${\bf E}^{\rm out}$  can be written as (see Ref.~\cite{karan2022ao} for the details of the calculations)
\begin{multline}\label{S0}
S^{\rm out}_0 = |E^{\rm out}_x|^2 +  |E^{\rm out}_y|^2\\
= \dfrac{1}{4}\left[ 2|T^{\rm s}_{\rm KM}|^2 + |T^{\rm p}_{\rm KM}|^2 + \cos 2\psi_{\rm in}\left\lbrace |T^{\rm p}_{\rm KM}|^2 - 2\left(|T^{\rm p}_{\rm KM}|^2  - |T^{\rm s}_{\rm KM}|^2\right)\cos 2\phi\right \rbrace + 2|T^{\rm p}_{\rm KM}|^2 \sin^2 \psi_{\rm in}  \right. \\
\left.-2 \left(|T^{\rm p}_{\rm KM}|^2
 - |T^{\rm s}_{\rm KM}|^2\right)\cos \delta_{\rm in} \sin 2\psi_{\rm in}\sin 2 \phi \right],
\end{multline}
\begin{multline}\label{S1}
S^{\rm out}_1 = |E^{\rm out}_x|^2 - |E^{\rm out}_y|^2\\
= \dfrac{1}{4}\left[2\left( |T^{\rm s}_{\rm KM}|^2 - |T^{\rm p}_{\rm KM}|^2\right)\cos 2\phi + \left\lbrace|T^{\rm s}_{\rm KM}|^2 + |T^{\rm p}_{\rm KM}|^2 + \left(|T^{\rm s}_{\rm KM}|^2 + |T^{\rm p}_{\rm KM}|^2 \right)\cos 4\phi \right. \right.\\
\left.\left. + 4{\rm Re}\left[T^{\rm s}_{\rm KM} T^{\rm p*}_{\rm KM}\right] \sin^2 2\phi\right\rbrace \cos 2\psi_{\rm in}  + \left \lbrace -4{\rm Im}\left[T^{\rm s}_{\rm KM} T^{\rm p*}_{\rm KM}\right]\sin\delta_{\rm in}\sin 2\phi + \left(|T^{\rm s}_{\rm KM}|^2 + |T^{\rm p}_{\rm KM}|^2 \right.\right.\right.\\
\left.\left. \left.- 2{\rm Re}\left[T^{\rm s}_{\rm KM} T^{\rm p*}_{\rm KM}\right]\right)\cos \delta_{\rm in}\sin 4\phi \right\rbrace \sin 2\psi_{\rm in} \right] ,
\end{multline}
\begin{multline}\label{S2}
S^{\rm out}_2= 2 \mbox{Re}\left[{E_x^{\rm out}}^{*} E^{\rm out}_y\right]\\
= \dfrac{1}{4}\left[ \left\lbrace |T^{\rm s}_{\rm KM}|^2 + |T^{\rm p}_{\rm KM}|^2+ 2 {\rm Re}\left[T^{\rm s}_{\rm KM} T^{\rm p*}_{\rm KM}\right]- \left(|T^{\rm s}_{\rm KM}|^2 + |T^{\rm p}_{\rm KM}|^2  - 2 {\rm Re}\left[T^{\rm s}_{\rm KM} T^{\rm p*}_{\rm KM}\right]\right)\cos 4\phi \right\rbrace  \right. \\
\left. \times \cos \delta_{\rm in}\sin 2\psi_{\rm in} + 4 {\rm Im}\left[T^{\rm s}_{\rm KM} T^{\rm p*}_{\rm KM} \right] \cos 2\phi \sin \delta_{\rm in}\sin 2 \psi_{in} + 2 \left(|T^{\rm s}_{\rm KM}|^2- |T^{\rm p}_{\rm KM}|^2 \right)\sin 2\phi \right. \\
\left. + \left(|T^{\rm s}_{\rm KM}|^2 + |T^{\rm p}_{\rm KM}|^2  - 2 {\rm Re}\left[T^{\rm s}_{\rm KM} T^{\rm p*}_{\rm KM}\right]\right)\cos 2\psi_{\rm in}\sin 4\phi  \right],
\end{multline}
\begin{multline}\label{S3}
S^{\rm out}_3= 2 \mbox{Im}\left[{E^{\rm out}_x}^{*} E^{\rm out}_y\right]\\
= \left({\rm Im}\left[T^{\rm s}_{\rm KM} T^{\rm p*}_{\rm KM}\right]\cos \delta_{\rm in} \cos 2\phi -  {\rm Re}\left[T^{\rm s}_{\rm KM} T^{\rm p*}_{\rm KM}\right] \sin \delta_{\rm in}\right)\sin 2\psi_{\rm in} \\
-  {\rm Im}\left[T^{\rm s}_{\rm KM} T^{\rm p*}_{\rm KM}\right]\cos 2\psi_{\rm in} \sin 2 \phi.
\end{multline}
\begin{figure}[t!]
\centering
\includegraphics[scale=0.9]{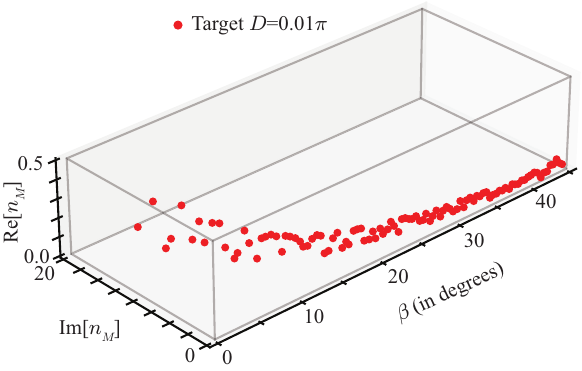}
\caption{Plot of the numerically optimized  ${\rm Re}[n_M]$ and ${\rm Im}[n_M]$ at 100 $\beta$ values for the target $D=0.01 \pi$ with a vertically polarized input field.}
\label{fig:plot_re_im_beta_at_d_1}
\end{figure}
\begin{figure}[t!]
\centering
\includegraphics[scale=0.75]{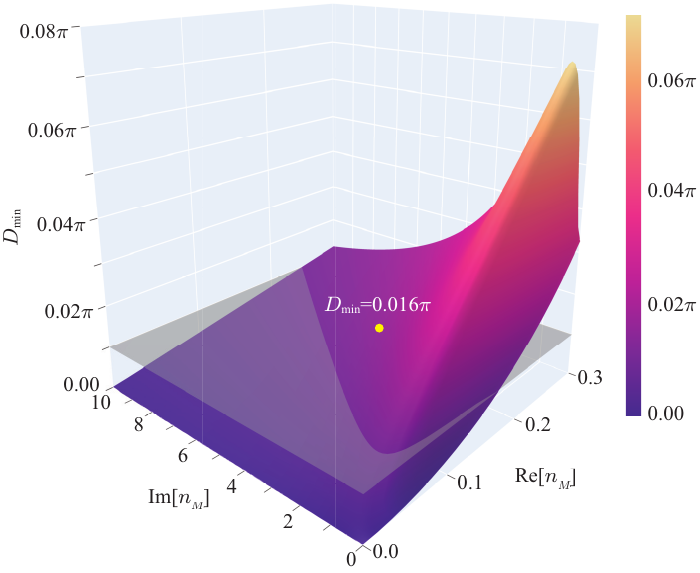}
\caption{The minimum mean polarization change  $D_{\rm min}$, obtained numerically by optimized over $\beta$ as a function of $n_M$. For a range of $n_M$ values, $D_{\rm min}$ is less than $0.01\pi$ while for the other $n_M$ values $D_{\rm min}$ is larger than $0.01\pi$. The yellow dot represents $D_{\rm min}$ corresponding to $n_M = 0.1568 + 3.8060 i$, the refractive index of the commonly available silver-coating. }
\label{fig:final_plot_re_im_D}
\end{figure}
The normalized Stokes parameters of the transmitted field are given by $\bar{S}^{\rm out}_0={S^{\rm out}_0}/{S^{\rm out}_0}$, $\bar{S}^{\rm out}_1={S^{\rm out}_1}/{S^{\rm out}_0}$, $\bar{S}^{\rm out}_2={S^{\rm out}_2}/{S^{\rm out}_0}$ and $\bar{S}^{\rm out}_3={S^{\rm out}_3}/{S^{\rm out}_0}$. The blue line in Fig~\ref{fig:schematic_of_km_and_poincare}(b) is a representative plot of the normalized Stokes parameters of the transmitted field as a function of the rotation angle from $\phi = 0^{\circ}$ to $\phi= 180^{\circ}$. The mean polarization change $D$ due to rotation is quantified as \cite{karan2022ao}:
\begin{equation}\label{mean_pol_change}
D = \frac{\int d~ ds}{\int ds} \\
=\dfrac{\int^{\pi}_{\phi=0} \cos^{-1} \left[ \sum^{3}_{i=1}\bar{S}^{\rm out}_i\bar{S}^{\rm in}_i\right] \sqrt{\sum^{3}_{i=1}\left(\dfrac{d \bar{S}^{\rm out}_i}{d\phi}\right)^2 }d\phi}{\int^{\pi}_{\phi=0} \sqrt{\sum^{3}_{i=1}\left(\dfrac{d \bar{S}^{\rm out}_i}{d\phi}\right)^2 }d\phi},
\end{equation}
where $d$ is the geodesic distance between $P_{\rm in}$ and the polarization state $P$ of the transmitted field at $\phi$, and $ds$ is the infinitesimal small arclength of the closed loop. The value of $D$ ranges from $0$ to $\pi$. $D=0$ represents no polarization change and $D=\pi$ represents maximum polarization change.

\begin{figure}[b!]
\centering
\includegraphics[scale=0.8]{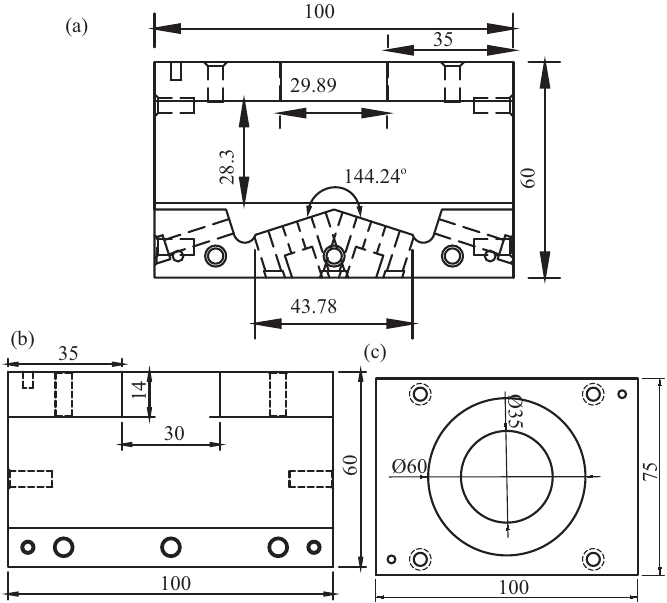}
\caption{Design and layout of the K-mirror. (a)  CAD drawing of the side-view of the housing. The base of the housing is inclined at an angle $\beta=17.88^{\circ}$ on both sides. (b) CAD drawing of another side-view of the housing. (c) CAD drawing of the top-view of the housing.  }
\label{fig:layout_k_mirror}
\end{figure}
\begin{figure}[b!]
\centering
\includegraphics[scale=0.92]{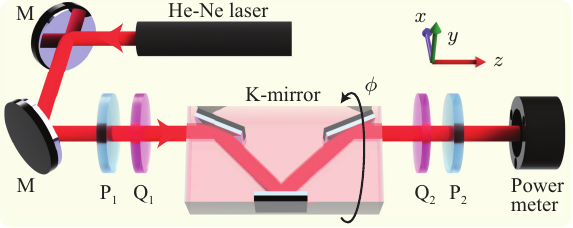}
\caption{Schematic of the experimental setup for measuring transmitted states of polarization using the home-built K-mirror at various rotation angles $\phi$. M : Mirror; $P_1$ and $P_2$ : Polarizers; $Q_1$ and $Q_2$ : Quarter-wave plates. }
\label{fig: exp_setup}
\end{figure}

In Ref.~\cite{karan2022ao}, $D$ was minimized by optimizing the base angle $\beta$ for a given value of the refractive index $n_M$ at three different states of incident polarization. In contrast, in this article, we present an optimization technique that is independent of the state of polarization. To this end, we define the quantity $F$ as 
\begin{equation}
F= |T^{s}_{\rm KM}- T^{p}_{\rm KM}|,
\end{equation}
which is a measure of the distance between the $s-$ and $p-$ component of the transfer matrix shown in Eqn.(\ref{eqn: transfer_mtrix}). We note that in the limit when $F$ tends to zero, $T_{\rm KM}\left(\phi\right)$ becomes an identity matrix $\mathbb{I}_{2 \times 2}$ and thus becomes independent of $\phi$.  The values of $n_M$ and $\beta$ that minimize $F$ also minimize $D$ independent of the state of the incident polarization. 

\begin{figure*}[t!]
\centering
\includegraphics[scale=0.765]{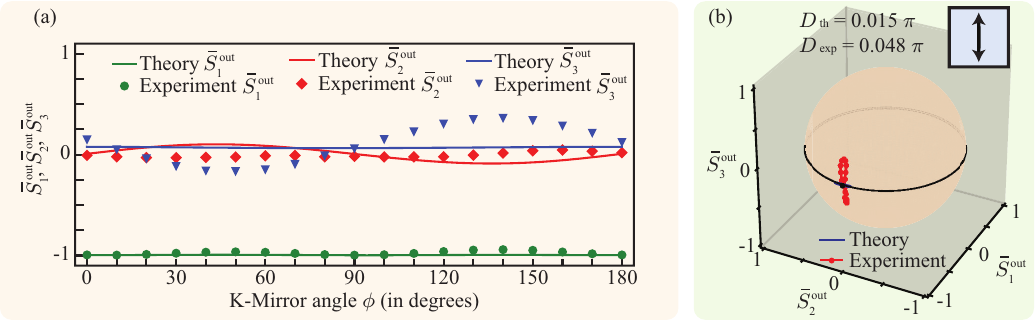}
\caption{ Experimentally measured normalized Stokes parameters of the transmitted field for a vertically polarized incident field. (a) Plots of $\bar{S}_1^{\rm out}$,  $\bar{S}_2^{\rm out}$, and  $\bar{S}_3^{\rm out}$ as a function of $\phi$. (b) \Poincare sphere representation of the transmitted field as a function of $\phi$.}
\label{fig: exp_th_result_vertical}
\end{figure*}
\begin{figure*}[t!]
\centering
\includegraphics[scale=0.765]{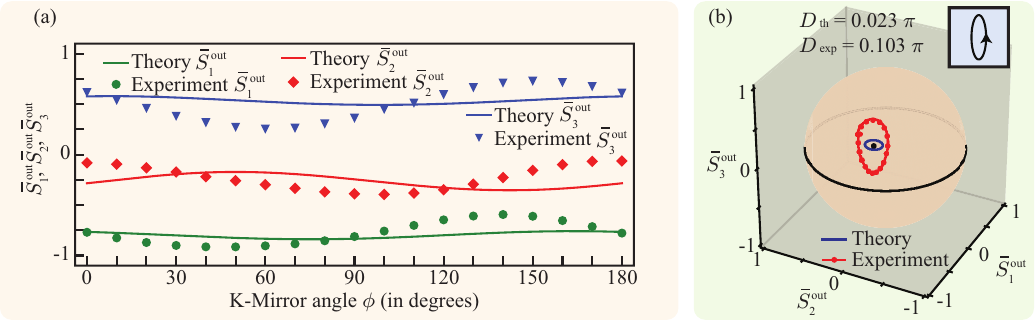}
\caption{Experimental plots of normalized Stokes parameters of the transmitted field for an elliptically polarized incident field. (a) Plots of $\bar{S}_1^{\rm out}$,  $\bar{S}_2^{\rm out}$, and  $\bar{S}_3^{\rm out}$ as a function of $\phi$. (b) \Poincare sphere representation of the transmitted field as a function of $\phi$.}
\label{fig: exp_th_result_elliptical}
\end{figure*}
\begin{figure*}[t!]
\centering
\includegraphics[scale=0.765]{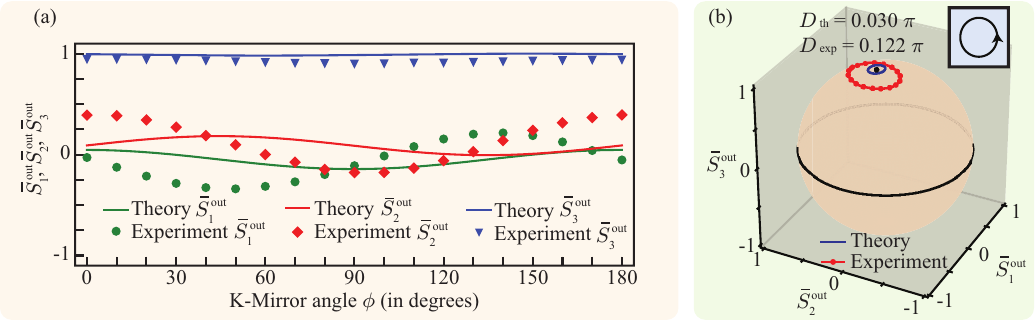}
\caption{Experimentally measured normalized Stokes parameters of the transmitted field for a circularly polarized incident field. (a) Plots of $\bar{S}_1^{\rm out}$,  $\bar{S}_2^{\rm out}$, and  $\bar{S}_3^{\rm out}$ as a function of $\phi$. (b) \Poincare sphere representation of the transmitted field as a function of $\phi$.}
\label{fig: exp_th_result_circular}
\end{figure*}

Although it is desirable to optimize $\beta$ and $n_M$ such that $F$ as well as $D$ are close to zero, it gets increasingly more difficult to practically design the optimized refractive index $n_M$ for target $D$ closer to zero. In such cases, one could optimize for near-zero target $D$ value. So, next, we present our numerical calculations to show that it is in principle possible to optimize $n_M$ for any given target $D$ and base angle $\beta$. Figure \ref{fig:plot_re_im_beta_at_d_1} represents the plot of the numerically optimized  ${\rm Re}[n_M]$ and ${\rm Im}[n_M]$ at 100 $\beta$ values for the target $D=0.01 \pi$ with a vertically polarized input field. For the results shown in Fig.~\ref{fig:plot_re_im_beta_at_d_1}, the optimization has been carried out in the following manner. First, we choose a value of $\beta$ and then we create a meshgrid of ${\rm Re}\left[n_M\right]$ and  ${\rm Im}\left[n_M\right]$. We first work with a coarse meshgrid of ${\rm Re}\left[n_M\right]$ and  ${\rm Im}\left[n_M\right]$ to have a rough estimate of the range of ${\rm Re}\left[n_M\right]$ and  ${\rm Im}\left[n_M\right]$ and then optimize with a finer meshgrid. For the results shown in Fig.~\ref{fig:plot_re_im_beta_at_d_1}, we take the finer meshgrid of dimensions $400\times 2500$, with  ${\rm Re}\left[n_M\right]$ ranging from $0$ to $0.4$  and ${\rm Im}\left[n_M\right]$ ranging from $0$ to $20$. We calculate $D$ for all possible pair of values of ${\rm Re}\left[n_M\right]$ and ${\rm Im}\left[n_M\right]$ and keep the pair that gives $D$ closest to $0.01\pi$. 
\begin{figure*}[t!]
\centering
\includegraphics[scale=0.76]{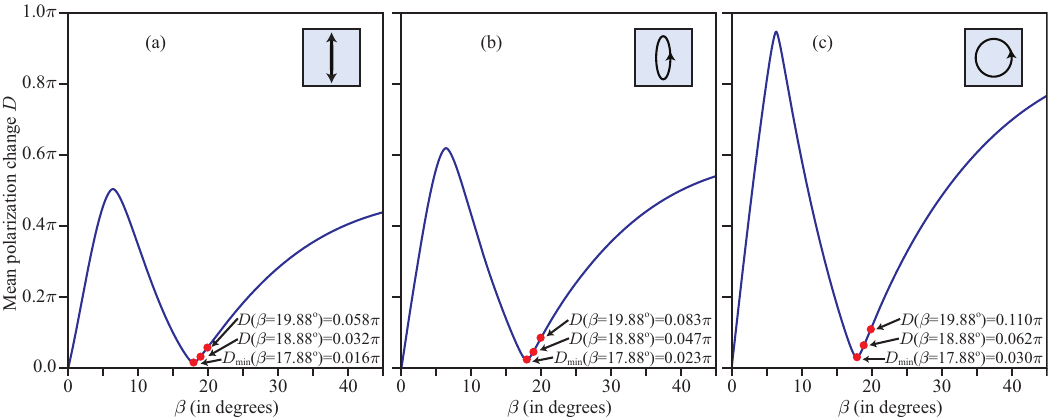}
\caption{{Dependence of $D$ on the base angle $\beta$ at $n_M = 0.1568 + 3.8060 i$ for three different states of polarization.} (a) incident field with vertical polarization, (b) incident field with elliptical polarization, and (c) incident field with circular polarization.}
\label{fig:D_vs_beta_for_three_incident_state_pol}\end{figure*}

This way, one can always find a physical $n_M$ for a given $\beta$ and target $D$. We note that using recently developed techniques \cite{meiers2023optexp, markel2016josaa}, it is in principle possible to design the reflective coating of the mirrors with any refractive index value. However, it gets increasingly more difficult to practically design $n_M$ for target $D$ very close to zero.

We note that while it is always possible to obtain a physical refractive index for a given $\beta$ and target $D$, it is in general not possible to find a physical $\beta$ for an arbitrary $n_M$ and target $D$. Figure \ref{fig:final_plot_re_im_D} shows minimum achievable mean polarization change $D_{\rm min}$ optimized over $\beta$ for a range of refractive indices $n_M$. For all the simulated points, the incident field has been kept vertically-polarized. We find that for a given $n_M$, $D_{\rm min}$ cannot go below a certain value. For illustrating this, we have plotted in Figure \ref{fig:final_plot_re_im_D} the floor at $D_{\rm min}=0.01\pi$. We note that for a range of $n_M$, $D_{\rm min}$ is less than $0.01\pi$ while for the other values, $D_{\rm min}$ is larger than $0.01\pi$. The yellow dot represents $D_{\rm min}$ corresponding to $n_M = 0.1568 + 3.8060 i$, the refractive index of the commonly available silver-coated mirror, in which case $D_{\rm min}=0.016 \pi$ and the corresponding optimized $\beta= 17.88^{\circ}$.

\section{Experiments}
We next report experimental demonstration of a K-mirror with refractive index $n_M = 0.1568 + 3.8060 i$ for which the optimized $\beta= 17.88^{\circ}$.  Figure \ref{fig:layout_k_mirror} shows the design and the layout of the housing of the K-mirror. We use three Thorlabs protected silver flat mirrors, each with $1$-inch diameter. As shown in Figs.~\ref{fig:schematic_of_km_and_poincare}(a) and \ref{fig:layout_k_mirror}, mirrors $M_1$ and $M_3$ are fixed on the base of the housing with tip-tilt adjustments. The housing is machine cut such that the base angle $\beta$ is equal to $17.88^{\circ}$ on both sides. The mirror $M_2$ is mounted on the top plate of the housing with a push-pull screw for precise height adjustments. The housing is rotated about an axis parallel to the optical table using a ball-bearing support. The dependence of the length $L$ and height $H$ on the base angle $\beta$ and clear aperture $h$ is given by: $L = 2 h~\mbox{cot}~ \beta$ and $H= (h/2)\left[1 + \mbox{tan}~2\beta / \mbox{tan}~\beta \right].$ \cite{karan2022ao}. The overall dimension of the housing is $L=130$ mm $\times$ $W=75$ mm  $\times$ $H=110$ mm. 

Figure~\ref{fig: exp_setup} depicts the experimental setup used for measuring the Stokes parameters as a function of $\phi$ and thus the mean polarization change $D$. We generate light fields with different states of polarization using a $5$ mW Newport He-Ne laser of wavelength $633$nm. For generating linearly polarized light, we keep polarizer $P_1$ in place, while for generating elliptically- and circularly-polarized light fields, we introduce a quarter-wave plate $Q_1$, immediately after $P_1$ with its fast axis rotated at $72^{\circ}$ and $45^{\circ}$ with respect to $-x$ direction, respectively. The generated field then goes through the K-mirror. The rotation axis of the k-mirror and the field propagation direction are made collinear by a very careful alignment procedure. This is done to ensure that the angular deviation of the transmitted beam is minimum. The Stokes parameter of the transmitted field as a function of $\phi$  is obtained using quarter-wave plate $Q_2$, polarizer $P_2$, and a Newport 1830-R optical power meter \cite{goldstein2017crc}.

Figures~ \ref{fig: exp_th_result_vertical}, \ref{fig: exp_th_result_elliptical}, and \ref{fig: exp_th_result_circular} present experimentally measured normalized Stokes parameters of the transmitted field at different rotation angles $\phi$ for vertically-, elliptically- and circularly-polarized incident fields, respectively. In each of the three figures, subfigure (a) plots the normalized Stokes parameters $\bar{S}_1^{\rm out}$, $\bar{S}_2^{\rm out}$, and $\bar{S}_3^{\rm out}$ as a function of $\phi$. The green, red, and blue solid-lines represent the theory, while the green dots, red diamonds, and blue downward triangles represent the experimentally measured values. The subfigure (b) in each of the three plots presents the  \Poincare sphere representation of the transmitted field as a function of $\phi$ ranging from $0^{\circ}$ to $180^{\circ}$. The blue line and the red dotted-curve denote the theory and the experimentally measured values. The black dot on the \Poincare sphere and the inset represent the polarization state of the input field.  The theoretically predicted mean polarization change $D_{\rm th}$ and the experimentally measured mean polarization change $D_{\rm exp}$ are indicated in each subfigure (b).

We note that the experimentally measured values of the mean polarization change $D_{\rm exp}$ are larger than the theoretical values $D_{\rm th}$. There are several reasons for this, including imperfect alignment, machining error, and the text-book value of the refractive index of the silver-coated mirror not being equal to refractive index of the mirrors used. Nonetheless, we find that the major source of the error is the machining error for our home-built K-mirror. Figure \ref{fig:D_vs_beta_for_three_incident_state_pol}(a), \ref{fig:D_vs_beta_for_three_incident_state_pol}(b), and \ref{fig:D_vs_beta_for_three_incident_state_pol}(c) show the dependence of $D$ on the base angle $\beta$ at $n_M = 0.1568 + 3.8060 i$ for vertical, elliptical, and circular incident polarizations, respectively. In each of the three plots, the corresponding $D$ values have been indicated in the figures at $\beta= 17.88^{\circ}$,  $\beta=(17.88+1)^{\circ}$, and $\beta=(17.88+2)^{\circ}$. We find that a machining error of $1^{\circ}$ in $\beta$ causes about a factor of 2 increase in $D$ from its minimum value $D_{\rm min}$ while a machining error of $2^{\circ}$ causes about a factor or 4 increase in $D$. Our K-mirror is home-built with local workshop facility, which had limited machining precisions. Nonetheless, such machining error can be easily overcome with commercial machining tools.

\section{Conclusion}
In conclusion, in this article, we generalize the results  reported in Ref.~\cite{karan2022ao} and use an optimization technique that is independent of the state of polarization of the incident field. We have shown that by optimizing the refractive index it is possible to design a K-mirror at any given base angle and with any value for the mean polarization change, including near-zero values. We have also designed and experimentally demonstrated a K-mirror with an order-of-magnitude lower mean polarization change than that of the commercially available K-mirrors. This can have important practical implications for OAM-based  applications that require precise wavefront rotation control without appreciable change in the state of polarization.

\begin{backmatter}
\bmsection{Funding}
Science and Engineering Research Board
with Grant No. STR/2021/000035 and  Grant No. CRG/2022/003070;
Department of Science \& Technology, Government
of India, with Grant No. DST/ICPS/QuST/Theme-1/2019

\bmsection{Acknowledgments}
S.K. thanks the University Grant Commission (UGC),
Government of India, for financial support. N.S. thanks the Prime Minister’s Research Fellowship (PMRF), Government of India. 

\bmsection{Disclosures} 
The authors declare no conflicts of interest.

\smallskip

\bmsection{Data availability} Data underlying the results presented in this paper are not publicly available at this time but may be obtained from the authors upon reasonable request.


\end{backmatter}

\bibliography{k_mirror_polarization_experiment_ref}

\end{document}